%% file: main.tex
\documentclass[10pt,twoside,letter]{scrartcl}

\input{packages}

\input{macros}

\input{author}

\begin{document}
\maketitle

\begin{abstract}
  \input{sections/0-abstract}
\end{abstract}

\begin{multicols}{2}
  \input{sections/1-introduction}
  \input{sections/2-methods-1}
  \input{sections/3-methods-2}
  \input{sections/4-results}
  \input{sections/5-applications}
  \input{sections/6-discussion}

  {\footnotesize \printbibliography}

\end{multicols}
\input{sections/si}
\end{document}

%% file: packages.tex
\usepackage[nonatbib]{styles/arxiv}

\usepackage[utf8]{inputenc} %
\usepackage[T1]{fontenc}    %
\usepackage[colorlinks]{hyperref}       %
\usepackage{url}            %
\usepackage{booktabs}       %
\usepackage{amsfonts}       %
\usepackage{nicefrac}       %
\usepackage{microtype}      %
\usepackage{multicol}
\usepackage{fancyhdr}
\usepackage{graphicx}
\usepackage{anyfontsize}
\usepackage{xcolor}
\usepackage{mathtools}

\usepackage{amsthm,thmtools}
\usepackage{styles/widetext}
\usepackage{bbm}
\usepackage{float}
\usepackage{cuted}

\usepackage{bookmark}
\usepackage{afterpage}

\usepackage{tikz-cd}

\hypersetup{
  citecolor=NavyBlue,
  linkcolor=NavyBlue,
  filecolor=NavyBlue,
  urlcolor=black
}

\usepackage[
backend=biber,
style=bibstyle,
articletitle=true,
isbn=false,
natbib
]{biblatex}

\newcommand{\orcid}[1]{\href{https://orcid.org/#1}{\hspace{5pt}\includegraphics[width=10pt]{figures/orcid.png}}}

\usepackage{academicons}
\definecolor{orcidlogocol}{HTML}{A6CE39}

\addtokomafont{footnote}{\small\footnotefont}

\usepackage[english]{babel}

\newcommand{\reffig}[2]{\hyperref[#1]{Figure}~\ref{#1}\hyperref[#1]{#2}}
\newcommand{\reftab}[1]{\hyperref[#1]{Table}~\ref{#1}}
\newcommand{\refvid}[1]{\hyperref[#1]{Video}~\ref{#1}}

%% file: macros.tex
\definecolor{darkblue}{RGB}{8, 60, 125}
\definecolor{darkgreen}{RGB}{46, 151, 78}
\definecolor{darkpurple}{RGB}{97, 64, 155}

%% file: author.tex
\date{\today} \title{Visual tracking brain computer interface}

\author{
  {Changxing Huang}\textsuperscript{1},
  {Nanlin Shi}\textsuperscript{1},
  {Yining Miao}\textsuperscript{1},
  {Xiaogang Chen}\textsuperscript{2},
  {Yijun Wang}\textsuperscript{3},
  {Xiaorong Gao}\textsuperscript{1,*}
  }  
\address{ \textsuperscript{1,*}Department of Biomedical Engineering, School of Medicine, Tsinghua University, Beijing, 100084, China. \\
  \textsuperscript{2}Institute of Biomedical Engineering, Chinese Academy of Medical Sciences and Peking Union Medical College, Tianjin, 300192, China.\\
  \textsuperscript{3}State Key Laboratory on Integrated Optoelectronics, Institute of Semiconductors, Chinese Academy of Sciences Beijing, 100083, China.\\
  \textsuperscript{*}email: gxr-dea@mail.tsinghua.edu.cn}

%% file: sections/0-abstract.tex
Brain-computer interfaces (BCIs) offer a way to interact with computers without relying on physical movements. Non-invasive electroencephalography (EEG)-based visual BCIs, known for efficient speed and calibration ease, face limitations in continuous tasks due to discrete stimulus design and decoding methods. To achieve continuous control, we implemented a novel spatial encoding stimulus paradigm and devised a corresponding projection method to enable continuous modulation of decoded velocity. Subsequently, we conducted experiments involving 17 participants and achieved Fitt’s ITR of 0.55 bps for the fixed tracking task and 0.37 bps for the random tracking task. The proposed BCI with a high Fitt's ITR was then integrated into two applications, including painting and gaming. In conclusion, this study proposed a visual BCI-based control method to go beyond discrete commands, allowing natural continuous control based on neural activity.

%% file: sections/1-introduction.tex
In the age of increasing demand for interaction with electronic devices, brain-computer interfaces (BCIs) offer a new approach that enables computer interaction without relying on muscular activity \supercite{gao2021interface}. Numerous studies have demonstrated the efficacy of BCIs in invasive experimental settings \supercite{pandarinath2017high,degenhart2020stabilization,willett2021high,anumanchipalli2019speech,metzger2023high,willett2023high,hochberg2006neuronal}. However, non-invasive BCIs based on electroencephalography (EEG) hold the potential for widespread use in everyday life. EEG-based BCIs typically generate commands by decoding motor intention activity or visual evoked responses\supercite{neuper2005imagery,pfurtscheller2001motor,sutter1992brain,bin2009vep}. Among these, 
visual BCIs, which decode visual responses, have gained popularity due to their efficient communication speed and minimal calibration requirements. They have been successfully applied 
to various interaction tasks such as typing \supercite{chen2015high,lin2006frequency,bin2009online,chen2021implementing}, drawing \supercite{munssinger2010brain,zickler2013brain}, cursor control \supercite{bi2014speed,allison2012hybrid,spuler2015brain}, and driving robotic devices \supercite{kishore2014comparison,muller2007control}.

Visual BCIs, despite their application in multiple scenarios, are often limited to classification tasks with discrete commands. The classical speller task has achieved an impressive 
Information Transfer Rate (ITR) of 325 bits per minute (bpm), allowing for the output of one of forty discrete commands in just 0.3 seconds \supercite{nakanishi2017enhancing}. However, these discrete commands are not suitable for continuous tasks such as drawing, cursor control, and robotic device control. The main challenge arises from the separation between the visual stimulus and the objects to be controlled. Consequently, users must constantly shift their attention between the visual stimulation and objects like cursor or robotic arms, creating a significant conflict in this setting. For example, in 2018, Chen et al. employed SSVEP-BCI to achieve robotic arm control, relying on step-by-step target selections \supercite{chen2018control}. However, this approach resulted in a significant time cost during the gaze shift process. In 2010, Muenssinger et al. developed a P300-Brain Painting application, which involved choosing elements and parameters for painting \supercite{munssinger2010brain}. Similarly, in 2015, Spüler designed a computer interaction system, where mouse movement was achieved through gaze shifts to select directional commands \supercite{spuler2015brain}. Both encountered the aforementioned challenge due to the necessity of multiple selections and attention shifts, resulting in inefficient and unnatural human-computer interactions. These studies indicate the inadequacy of discrete decoding methods for achieving continuous control and emphasize the need for innovative solutions within the traditional field of visual BCIs.

To enable natural and efficient interaction, motor imagery BCIs have successfully transformed discrete commands into continuous 2-D cursor control \supercite{silversmith2021plug,edelman2019noninvasive}. In 2004, Wolpaw et al. demonstrated the control of two-dimensional movement using noninvasive motor imagery BCIs \supercite{wolpaw2004control}, thereby highlighting that multidimensional control is not exclusive to invasive methods. In 2019, Edelman et al. introduced and validated a noninvasive framework using EEG to achieve the neural control of a robotic device for continuous random target tracking \supercite{edelman2019noninvasive}. This work further underscores the capability of non-invasive BCIs to attain continuous control. In the realm of visual BCIs, the pivotal step towards achieving multidimensional continuous control lies in shifting the approach from classifying different classes of patterns to predicting continuous variables from neural activities.

In this study, we introduce a novel visual tracking BCI to achieve continuous cursor control. In contrast to the classical BCI speller, our approach focuses on continuous movement tasks such as cursor control. During the tasks, participants gaze at the intended position under the multi-region stimulus paradigm while the cursor tracks the position using EEG predictions. To enable continuous decoding, we map the correlation coefficients of eight distinct patterns in the paradigm to their corresponding directions. Adding velocity weight that was trained using the least square method, a corrected velocity vector can be predicted to control cursor movement. Our proposed visual tracking BCI achieves a Fitt's ITR of 0.55 bps for fixed tracking tasks and 0.37 bps for random tracking tasks. Additionally, we showcase various applications, including painting and gaming, to demonstrate the versatility of this system. Overall, this study presents a significant advancement beyond the limitations of traditional discrete classification modes in visual BCIs, enabling natural and efficient continuous control based on neural activity.

%% file: sections/2-methods-1.tex
\begin{figure}[H]
  \centering
  \includegraphics[width=\columnwidth]{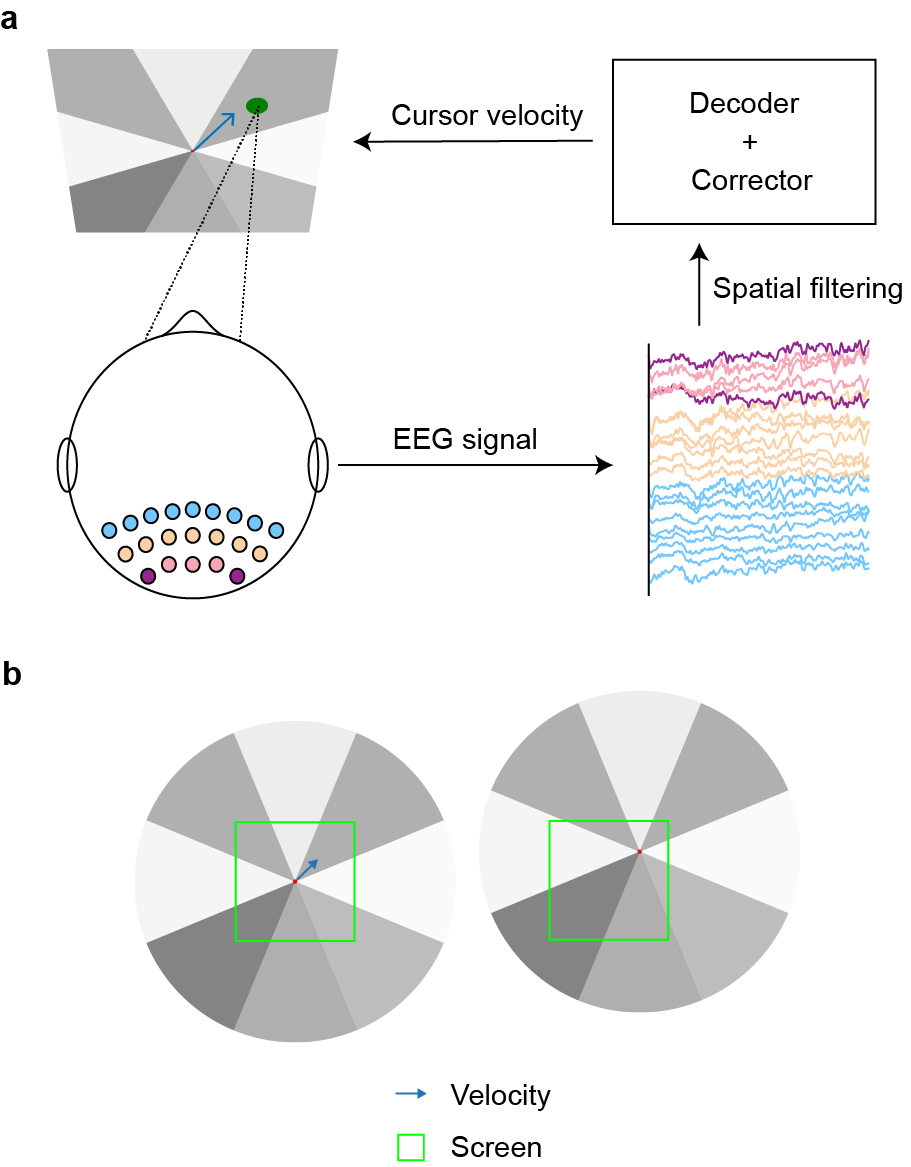}
  \caption{\textbf{ Visual tracking BCI system. a,} Schematic of the visual tracking BCI framework. EEG signals recorded from multiple electrodes are passed to the decoder after spatial filtering. The velocity predicted by the decoder is corrected by the corrector to guide cursor movement. \textbf{b,} Stimulus implementation. Cursor movement is achieved by updating and shifting the background image.
  }
\label{fig:1}
\end{figure}

\section{Methods}\label{sec:2}
\subsection{Stimulus paradigm}
\emph{Multi-region stimulus.} Different from classical speller design, the paradigm aims to track the location of fixation and fulfill cursor control based on neural activity. When fixating on the target location (green circle in \reffig{fig:1}{(\textbf{a})}), the cursor tracks and automatically moves to the location based on EEG activity. To achieve this, the whole visual field is separated into 8 radial stimuli, centering the cursor. The relative position between the stimulus and the cursor remains constant throughout the cursor movement. 

To achieve continuous control, each region of the stimulus is modulated by a sequence of uniformly-distributed White Noise (WN; see \reffig{fig:S1}{}). The WN-based modulation belongs to cVEP BCI and was recently proposed to exhibit multifold advantages over traditional sinusoidal coding \supercite{shi2023estimating}. The significant advantages of WN modulation are that the broadband tagging stimulus balances the spectral preference of the primary visual channel, thus avoiding uneven decoding results that prefer a certain class. To put it in the context of continuous control, expressing a preference for a particular radial stimulus could lead to potential directional drift and an increased probability of errors.

\emph{Stimulus implementation.} In the experiment, the screen was set to 800×800 pixels with a black background. On top of the background, the 8 stimulus regions flickered with 1 s of WN sequence and repeated consecutively. The red point, situated at the center of the radial stimulus with a radius of 5 pixels, served as the cursor. The green target was a semi-transparent circle with a radius of 40 pixels. To reduce visual fatigue, the contrast limit of each region was kept to gray (RGB=(127, 127, 127)). This implies that the RGB sequences for each region are obtained by multiplying the WN sequences by (127, 127, 127). To implement the tracking experimental design, the stimulus material (picture form) was generated larger than the screen size (\reffig{fig:1}{(\textbf{b})}), which means only part of the stimulus is visible on the screen. As each movement proceeded, the stimulus picture and its center were adjusted correspondingly. The entire BCI system, including stimulation, acquisition, and operation, was implemented through PsychoPy.

\subsection{Experiments materials and design}
\emph{Subjects and devices.} Seventeen individuals (6 females, 7 experienced) participated in the study. Each participant was asked to read and sign an informed consent form approved by the Institution Review Board of Tsinghua University before the experiments.

The stimulus was displayed on a 1920×1080 ASUS MG279Q monitor with a 60 Hz refresh rate. EEG data were recorded by Synamps2 system (NeuroScan, Inc.) at a 1000 Hz sampling rate with 62 channels, and 21 of them (Pz, P1/2, P3/4, P5/6, P7/8, POz, PO3/4, PO5/6, PO7/8, Oz, O1/2, and CB1/2) were used for analysis. The recorded EEG data were first down-sampled to 250 Hz, then band-pass filtered to 4-100 Hz, and notch-filtered for the 50 Hz interference.

\emph{Experiment protocol.} The experiments contain three parts: training (Stage I, Stage II), testing (fixed tracking task, random tracking task), and application experience (painting, snake game). During the training phase, each participant was instructed to fixate on the green target in both Stage I and Stage II (\reffig{fig:2}{(\textbf{a})}), without receiving online feedback on cursor movement. The purpose of Stage I is to train a template matching model for calculating correlation coefficient matrix ($\boldsymbol{P}$; $\boldsymbol{P} \in \mathbb{R}^{1 \times N_r}$, $N_r$ stands for the number of stimulus regions), and $\boldsymbol{P}$ will be employed to predict velocity. In this stage, there were 8 targets, each located on the center line of its respective region, positioned at a distance of 1/3 of the screen width ($w_s$) from the center red point. The aim of Stage II is to train the corrected velocity weight for generating more efficient velocity for continuous cursor control (see Decoding algorithms for Velocity weight). To achieve this, we introduced 32 targets in this stage (as in \reffig{fig:2}{(\textbf{a})}), including intersected directions at the boundary between two adjacent regions and a new distance of 1/6 of $w_s$ from the center point (forming inner and outer circle). In both stages, each target was presented 6 times for a duration of 1 s each time.

During the testing phase, while participants fixated on the cued target, the EEG data collected at each step, with a duration of 1 s and no overlap, was used to determine the decoded velocity, which in turn controlled the cursor's movement towards the target. Then, the red cursor moved at the decoded velocity until it reached the target (\reffig{fig:2}{(\textbf{a})}). The testing phase contains two tasks: fixed tracking task and random tracking task. During the fixed tracking task, the green targets appear at the same 32 positions as in training Stage II. Participants are instructed to fixate on the cued target until the cursor hits the target, during which eye blinking is allowed. If the target is successfully reached within 15 s, the system provides feedback on the time cost and waits for the participant to press space to proceed to the next trial. If not, the system displays a timeout message and moves on to the next trial. At the beginning of each trial, the cursor is refreshed to the center of the screen. The fixed tracking task consists of 3 blocks, with the 32 positions repeated in each block. In the random tracking task, the green target appears at a random position on the screen. At the beginning of each consecutive trial, the cursor starts from the end position of the previous trial. In this task, a total of 12 trials are performed.

During application experience, participants were engaged in using the built-in applications, specifically painting and snake game. In the painting stage, participants were instructed to use their visual gaze as brushstrokes, allowing them to paint at their will. In the gaming stage, we used the downgraded version of continuous decoding, where the screen interface is split into blocks and the cursor moves one grid at a time. The cursor of this game is simplified as the head of the snake, and target is shown as the food. Once the snake obtains the food, its body length increases by the length of one block. Participants were instructed to increase the body length while avoiding collisions with the screen edges or the snake’s own body. Following the game, participants were awarded a score that corresponded to the snake’s length. The decoding algorithm of the snake game employs a dynamic stopping strategy based on confidence estimation (see Decoding algorithms for Confidence estimation based on hypothesis testing).

%% file: sections/3-methods-2.tex
\begin{figure*}[t!]
  \centering
  \includegraphics[width=\textwidth]{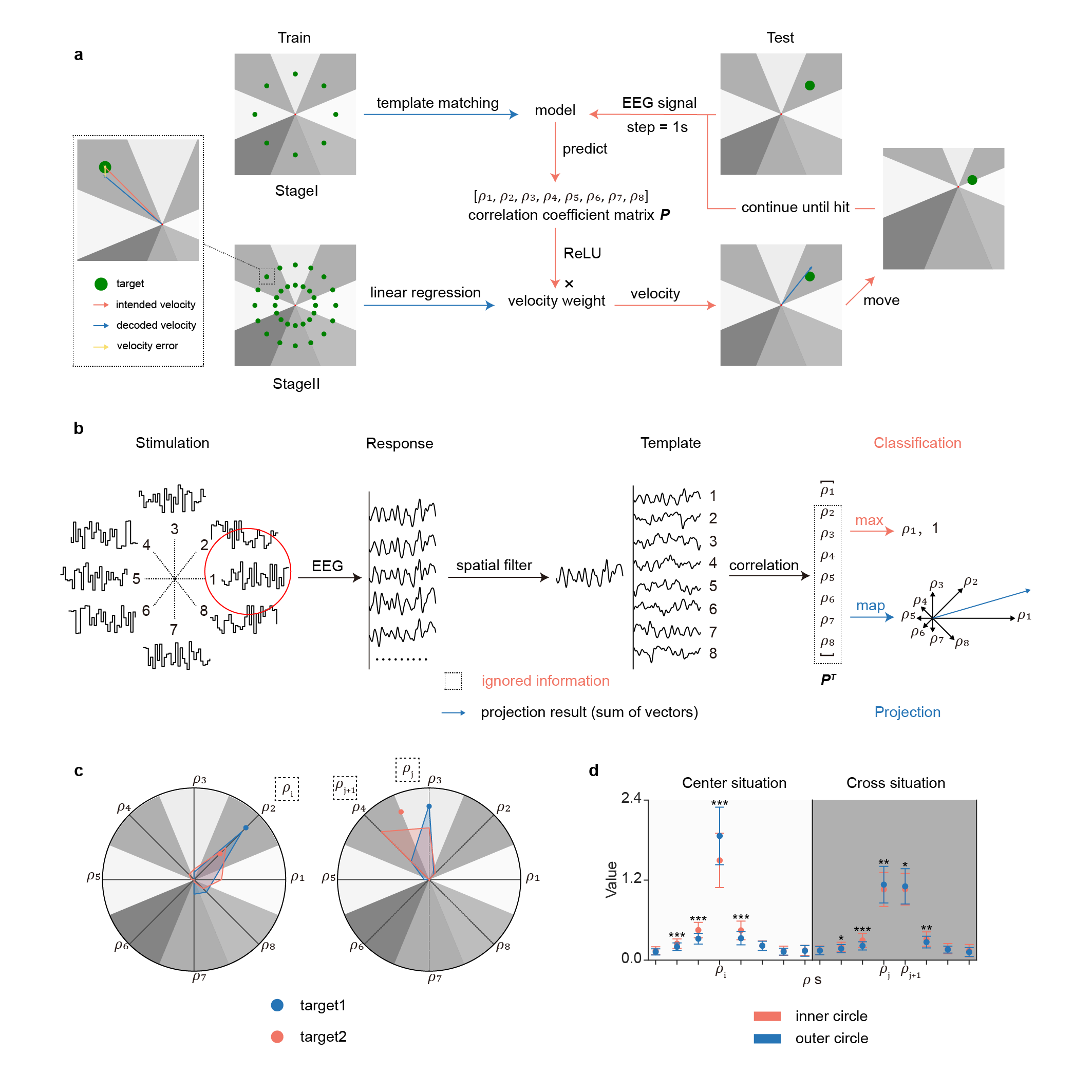}
  \caption{\textbf{Decoder design and feasibility. a,} The main experiment consists of training and testing parts. The training part is divided into stage I (for training template matching model) and stage II (for training velocity weight by linear regression). In the testing part, cursor moves at the decoded velocity until it hits the target. 
    \textbf{b,} Diagram of the decoding method. Projection method projects correlation coefficients onto the corresponding directions, utilizing information that is ignored by the classification method. \textbf{c,} Radar plot of the correlation coefficient distribution. Left: the examples of distribution for inner circle and outer circle positions; Right: the examples of distribution for center and cross situations. 
    \textbf{d,} Comparison of $\rho$s’ value in different conditions. Bars indicate mean ± s.d. Paired t-test (n = 17): *P < 0.05, **P < 0.01, ***P < 0.001.}
\label{fig:2}
\end{figure*}

\subsection{Decoding algorithms}
\emph{Spatial filtering.} We chose Task-Related Component Analysis (TRCA) as the template matching method \supercite{nakanishi2017enhancing}. 
The goal of TRCA is to find a set of spatial filters that maximize the task-related components by optimizing the inter-trial consistency. To achieve this, the TRCA algorithm maximizes the within-trial covariance and minimizes the between-trial covariance. The covariance within trials is calculated as:
\begin{equation}
\boldsymbol{w}^T \boldsymbol{S} \boldsymbol{w}=\sum_{\substack{h_1, h_2=1 \\ h_1 \neq h_2}}^{N_t} \sum_{j_1, j_2=1}^{N_c} w_{j_1} w_{j_2} \operatorname{Cov}\left(x_{j_1}^{\left(h_1\right)}(t), x_{j_2}^{\left(h_2\right)}(t)\right) 
\nonumber
\end{equation}
where $h$ means different trials and $j$ means different channels. $N_t$ is the number of trials, and $N_c$ is the number of channels. The between-trial covariance is calculated as:
\begin{equation}
\boldsymbol{w}^T \boldsymbol{Q} \boldsymbol{w}=\sum_{j_1, j_2=1}^{N_c} w_{j_1} w_{j_2} \operatorname{Cov}\left(x_{j_1}(t), x_{j_2}(t)\right)
\nonumber
\end{equation}
the final spatial filter can be determined as follows:
\begin{equation}
\widehat{\boldsymbol{w}}=\underset{\boldsymbol{w}}{\arg \max } \frac{\boldsymbol{w}^T \boldsymbol{S} \boldsymbol{w}}{\boldsymbol{w}^T \boldsymbol{Q} \boldsymbol{w}}
\nonumber
\end{equation}
In the typical classification task, template matching is made by comparing the correlation coefficient matrix between spatial weighted evoked template and single trial response (\reffig{fig:2}{(\textbf{b})}). And the final prediction is made by selecting the stimulation class corresponding to the maximum correlation coefficient ($\rho_{max}$). It's important to note that, in the TRCA method, due to the use of filter banks, the final value of each correlation coefficient ($\rho$) is obtained by weighting the correlation coefficients (using $r$ as a distinct reference) between the sub-filtered response and the template. Consequently, the values of $\rho$s do not conform to the standard range of -1 to 1.
\begin{equation}
\rho=\sum_{m=1}^{N_m} a(m) \cdot r_m
\nonumber
\end{equation}
where $m$ means different sub-bands and $N_m$ is the total number of sub-bands. $a(m)$ is defined as $a(m) = m^{-1.25} + 0.25$. In practice, other template matching methods such as Task Discriminate Component Analysis (TDCA) or Filter Bank Canonical Correlation Analysis (FBCCA) can be chosen based on respective stimulus paradigms and training requirements.

\emph{Velocity prediction.} The key distinction from discrete classification to continuous tracking is to change template matching based on $\rho_{max}$ to coefficient projection based on the entire coefficients (\reffig{fig:2}{(\textbf{b})}). To produce continuous movement, we map the correlation coefficients ($\rho$s) of distinct patterns to $\rho$ vectors starting from the cursor center and point to their corresponding 8 directions. The sum of all $\rho$ vectors, as we call projection result in this study, represents the intended direction of cursor movement. The change from $\rho_{max}$ to $\rho$s promises a significant improvement of decoding algorithm. By leveraging the correlation coefficients from other stimulus regions, we quantified the combined effects of various stimulus regions in the visual field. It’s worth noting that the correlation coefficients can take on negative values, and the aforementioned projection method conducts unidirectional projection. Consequently, ReLU is employed to reset negative values to zero. However, the projection result doesn’t align with practice scenarios, and due to variations in stimulus responses among participants, it lacks universality and may cause direction drift. To minimize the projection error and thrive for ideal tracking performance, we introduced the velocity weight for correction.

\emph{Velocity weight.} The velocity weight serves two main purposes. One is to ensure that the projection result corresponds to a real velocity that matches the screen size, while the other is to individually adjust and correct the magnitude and direction of the decoded velocity for each participant. The velocity vector ($\boldsymbol{v}$) can be calculated as the dot product of the correlation coefficient matrix ($\boldsymbol{P}$) and velocity weight ($\boldsymbol{V}_{\boldsymbol{w}}$; $\boldsymbol{V}_{\boldsymbol{w}} \in \mathbb{R}^{N_r \times 2}$, 2 represents the Cartesian coordinates x and y). 
\begin{equation}
\boldsymbol{v}^T=\boldsymbol{P} \cdot \boldsymbol{V}_{\boldsymbol{w}}
\nonumber
\end{equation}
We can align the decoded velocity with the screen size by using an initial velocity weight ($\boldsymbol{V}_{\boldsymbol{w}}^{init}$) distributed in a circular pattern with a radius of 1/3 of the screen width ($w_s$) divided by 2 (the average length of projection result).
\begin{equation}
  \boldsymbol{V}_{\boldsymbol{w}}^{init}=\left(\begin{array}{cc}
  1 & 0 \\
  \sqrt{2} / 2 & \sqrt{2} / 2 \\
  0 & 1 \\
  -\sqrt{2} / 2 & \sqrt{2} / 2 \\
  -1 & 0 \\
  -\sqrt{2} / 2 & -\sqrt{2} / 2 \\
  0 & -1 \\
  \sqrt{2} / 2 & -\sqrt{2} / 2
  \end{array}\right) \cdot \frac{w_s}{6}
  \nonumber
\end{equation}
To facilitate a more natural effect, during the practical experiment, the velocity gradually diminishes to zero within a time step (1 s). The average velocity of this process corresponds to the decoded velocity.

Furthermore, to minimize prediction error between intended velocity and decoded velocity (\reffig{fig:2}{(\textbf{a})}) and solve the issue of subject differences, we trained a corrected velocity weight ($\boldsymbol{V}_{\boldsymbol{w}}^{*}$) by least square method for each participant, using data from training Stage II:
\begin{equation}
\boldsymbol{V}_{\boldsymbol{w}}^{*}={(\boldsymbol{D}^T\boldsymbol{D})}^{-1}\boldsymbol{D}^T\boldsymbol{I}
\nonumber
\end{equation}
where $\boldsymbol{D}$ is the decoded correlation coefficient matrix for all trials ($\boldsymbol{D} \in \mathbb{R}^{N_t \times N_r}$, $N_t$ is the number of trials and $N_r$ is the number of stimulus regions), and $\boldsymbol{I}$ is the intended velocity matrix for all trials ($\boldsymbol{I} \in \mathbb{R}^{N_t \times 2}$), where intended velocity is calculated as the distance from the target to the center red point divided by one time step (1 s). If there is no special prompt, the velocity weight in Results refers to the corrected (or trained) one.

\emph{Confidence estimation based on hypothesis testing.} In the snake game, where the snake advances one grid at a time using classification methods, we have integrated a dynamic stopping strategy. This strategy helps determine whether the snake should move or not, aiming to minimize the likelihood of game failures caused by incorrect judgments. The strategy evaluates prediction confidence based on the hypothesis testing framework that has been used in the previous studies. Based on the correlation coefficient matrix above, we calculate the confidence level between the maximum correlation coefficient and the remaining correlation coefficients. We can decide whether to reject the hypothesis and, accordingly, determine whether the snake should move by setting a threshold for the p-value.

\subsection{BCI performance evaluation}
Different from typical BCI speller which use ITR for evaluation metrics, we employ Fitt’s ITR to properly evaluate BCI performance in the continuous task \supercite{felton2009evaluation}. The Fitt’s ITR is related to three parameters, including the target size, distance to the target, and the time cost.
\begin{equation}
  \text { Fitt's ITR }=\frac{\log _2((D+S) / S)}{\text {Time }}
  \nonumber
\end{equation}
where $S$ represents the target size, $D$ represents the distance to the target, and $T$ represents the time cost (see \reffig{fig:S2}{}). The success rate is computed as the percentage of trials in which the cursor reaches the target within the time limit. The time to target is the time required for the cursor to move from the center or start position to the target. The unsuccessful trials are not involved in the calculation of Fitt’s ITR and time to target, while they are presented as success rate.

Other than Fitt’s ITR, we defined two errors as the metrics of the velocity prediction (\reffig{fig:3}{(\textbf{c})}). One is angular error, which measures the angular difference between the decoded velocity and the intended velocity. The other is vector error, which is calculated using the length of the difference between the intended velocity vector and the decoded velocity vector to divide the length of the intended velocity vector. The error calculation is based on the decoded velocity from the first step of each trial in the fixed tracking task. By combining Fitt’s ITR and velocity errors, we can yield objective measurements for the decoding algorithm and the resulting tracking performance.

%% file: sections/4-results.tex
\begin{figure*}[b!]
  \centering
  \includegraphics[width=\textwidth]{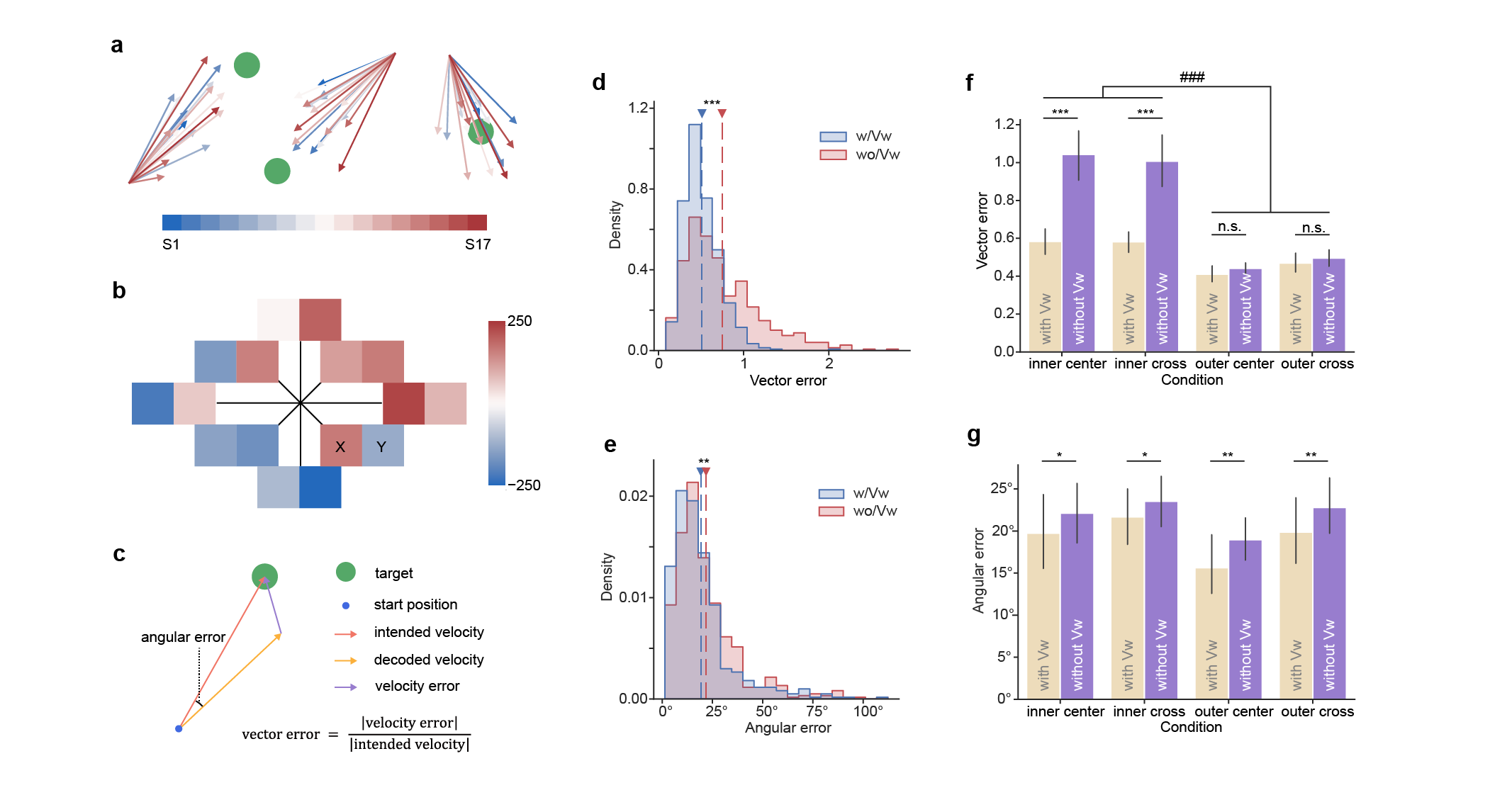}
  \caption{\textbf{Velocity prediction and velocity weight. a,} Decoded velocity of all participants 
    in three typical target positions. \textbf{b,} Velocity weight averaged for all participants (n = 17), 
    conforming to the form of the Cartesian coordinate system. \textbf{c,} Schematic diagram of vector error 
    and angular error. \textbf{d,} Distribution of vector error for all participants using corrected velocity 
    weight (w/Vw) and using only the initial velocity weight (wo/Vw) (paired t-test: P = 4.396E-5, n = 17; w/Vw: 
    mean = 0.51, s.d. = 0.08; wo/Vw: mean = 0.75, s.d. = 0.14). \textbf{e,} Distribution of angular error for all 
    participants (paired t-test: P = 0.0016, n = 17; w/Vw: mean = 19.1°, s.d. = 7.2°; wo/Vw: mean = 21.7°, s.d. = 6.0°). 
    \textbf{f,} Vector error of different conditions (‘inner center’: at the inner circle and the center of 
    the region, ‘inner cross’: at the inner circle and the boundary between the regions, so as for ‘outer center’ 
    and ‘outer cross’). \textbf{g,} Angular error of different conditions. Bars indicate mean ± s.d. Statistical 
    analysis using two-way ANOVA (\textbf{f}, n = 34 per group, $^{\#\#\#}$P < 0.001) and paired t-test (\textbf{f} and \textbf{g}, n = 17, *P < 0.5, **P < 0.01, ***P < 0.001).}
\label{fig:3}
\end{figure*}

\begin{figure*}[t!]
  \centering
  \includegraphics[width=\textwidth]{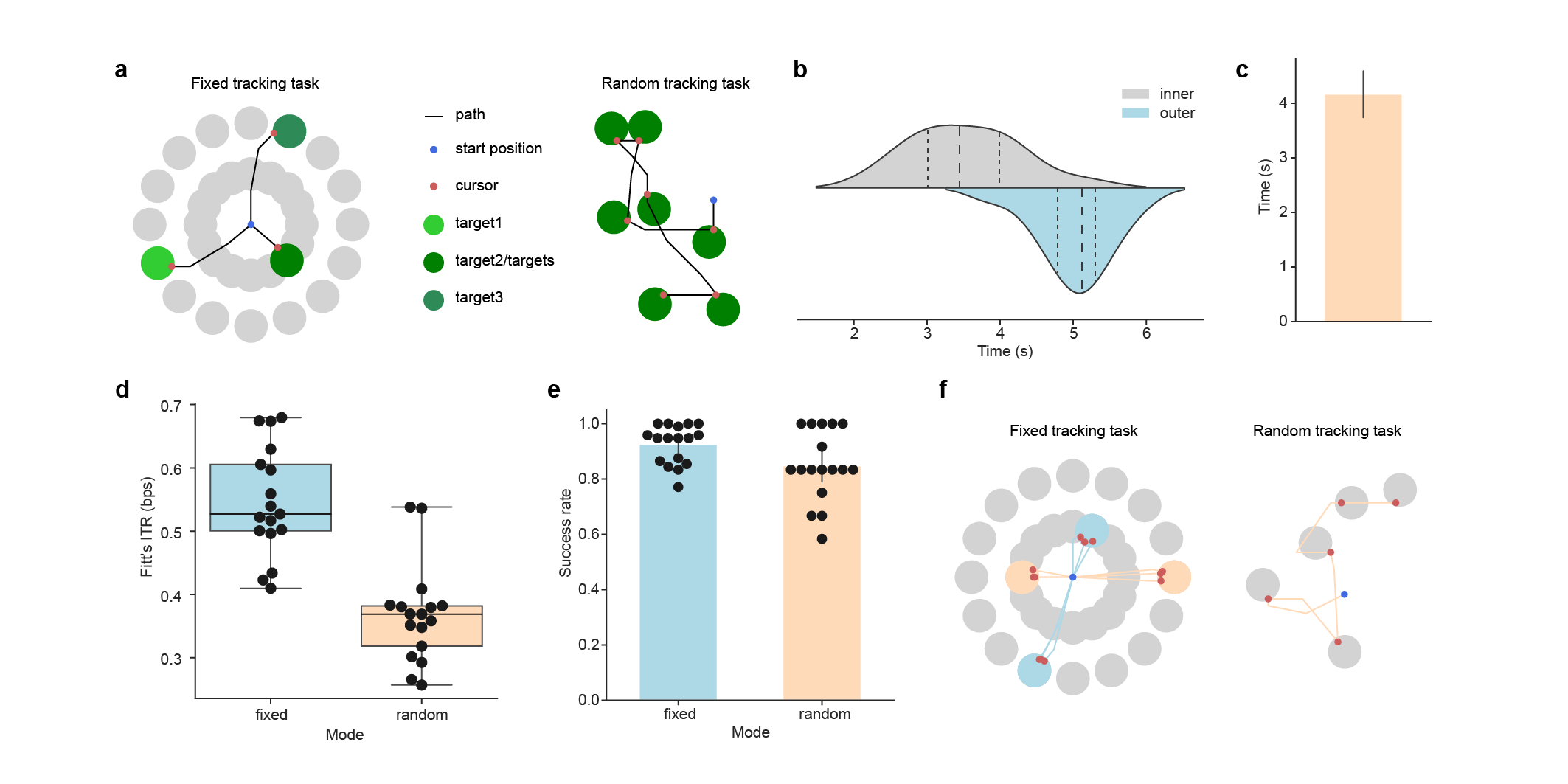}
  \caption{\textbf{Visual tracking task performance. a,} Schematic of fixed and random tracking tasks. \textbf{b,} Distribution of time to 
    target for fixed tracking task in inner circle and outer circle conditions. Dashed lines are quarters. \textbf{c,} Time to target for the random tracking task. 
    Bars indicate mean ± 95\% CI. \textbf{d,} Fitt’s ITR of fixed and random tracking tasks (n = 17, medians, interquartile ranges, maxima and minima). \textbf{e,} 
    Success rate of fixed and random tracking tasks (n = 17, bars indicate mean ± 95\% CI). 
    \textbf{f,} Tracking trajectories of fixed and random tracking tasks (four typical positions for the fixed tracking task, and the first 5 targets for the random tracking task).}
\label{fig:4}
\end{figure*}

\section{Results} \label{sec:3}
\subsection{Feasibility of projection method}
We observed that the distribution of correlation coefficients ($\rho$s) reflects the location information of users’ visual fixation point (\reffig{fig:2}{(\textbf{c})}). For the 32 target positions in training Stage II, when the target is located on the center line of the region (referred to as ‘center situation’), the $\rho_i$ for the corresponding region is significantly larger than the remaining regions (P < 0.001, unpaired t-test). Additionally, when comparing the effects of the inner and outer circles, we found that the $\rho_i$ of the outer circle is significantly larger than that of the inner circle (P = $2.85 \times 10^{-9}$, paired t-test), whereas other adjacent $\rho$s exhibit opposite trends (\reffig{fig:2}{(\textbf{d})}), indicating $\rho$s’ distribution can effectively align with the target direction and correspond to the target distance. In the case of the target located at the boundary between two adjacent regions (cross situation), both $\rho_j$ and $\rho_{j+1}$ (counter-clockwise ordering) corresponding to the two adjacent regions are relatively large, effectively reflecting the orientation of the target. The differences in $\rho$s between the inner circle and outer circle exhibit similarities to the center situation (P = $6.77 \times 10^{-3}$ for $\rho_j$, P = 0.0273 for $\rho_{j+1}$, paired t-test). In summary, the $\rho$s’ distribution can reflect the location information for any target positions. Based on this, a velocity that is able to automatically adjust both magnitude and direction can be derived. 

The mechanism through which this method functions is that human vision has receptive fields, and the neural activity is jointly encoded by stimulation from all regions. As shown in \reffig{fig:S3}{}, when fixating on nearby targets (in relation to the cursor), the higher proportion of surrounding stimulus within the field of view exerts a more pronounced influence, causing a dispersed distribution of $\rho$s towards the neighboring regions (in relation to the target). Conversely, when fixating on distant targets, the reduced proportion of surrounding stimulus within the field of view results in a more centralized distribution of $\rho$s towards the central regions (refer to \reffig{fig:2}{(\textbf{c-d})}). Here, we have generated the distribution of $\rho$s which can represent the users’ fixation location by employing the multi-region stimulus paradigm. By subsequently incorporating the corresponding decoding algorithm, we are able to generate the desired continuous variable, which is the velocity.

\subsection{Effect of velocity prediction and velocity weight}
We found that the decoding algorithm can effectively estimate the intended velocity, and the utilization of corrected velocity weight significantly minimizes both angular error and vector error. \reffig{fig:3}{(\textbf{a})} illustrates the decoded velocity of all participants in three typical target positions. All velocities are able to guide the cursor to approach the target, and in the case of certain subjects, the velocity accurately points to the target with the appropriate size and direction. In order to observe the characteristics of velocity weight, we presented the average corrected velocity weight for all participants corresponding to the projection direction in \reffig{fig:3}{(\textbf{b})}. As anticipated, the positive and negative relationship of velocity weight conforms to the Cartesian coordinate system, suggesting that the decoding algorithm leverages spatial encoding information from the stimulus paradigm. Furthermore, the selected stimulus sequences exhibit consistency within the population with no significant directional deviation (aligning with the characteristics of the WN sequence). The velocity weight for each participant was displayed in \reffig{fig:S4}{}.

We then compared the vector error and angular error of the decoded velocity between using the corrected velocity weight (w/Vw) and using only the initial velocity weight (wo/Vw), and found that the use of corrected velocity weight significantly reduced the vector error (\reffig{fig:3}{(\textbf{d})}; paired t-test: P = $4.396 \times 10^{-5}$, n = 17; w/Vw: mean = 0.51, s.d. = 0.08; wo/Vw: mean = 0.75, s.d. = 0.14) and angular error (\reffig{fig:3}{(\textbf{e})}; paired t-test: P = 0.0016, n = 17; w/Vw: mean = 19.1°, s.d. = 7.2°; wo/Vw: mean = 21.7°, s.d. = 6.0°) of the decoded velocity. Moreover, we examined the impact of target location (inner/outer, cross/center) on vector error and angular error. A significant improvement of vector error was noted in ‘inner center’ and ‘inner cross’ conditions (\reffig{fig:3}{(\textbf{f})}; P < 0.001, paired t-test), while no significant improvement was noted for ‘outer center’ and ‘outer cross’ conditions. This demonstrated that the corrected velocity weight mainly rectified the vector error of the inner circle position, which aligns with the finding that the vector error of the inner circle position was significantly larger than that of the outer circle position (F = 102.2, P = $3.53 \times 10^{-18}$, two-way ANOVA). In terms of angular error, a significant improvement was observed in all conditions. (‘inner center’: P < 0.05, ‘inner cross’: P < 0.05, ‘outer center’: P < 0.01, ‘outer cross’: P < 0.01). In summary, we discovered that the corrected velocity weight can significantly reduce both angular error and vector error. Particularly, the impact on vector error is most pronounced in the inner circle situation.

\begin{figure*}[b!]
  \centering
  \includegraphics[width=\textwidth]{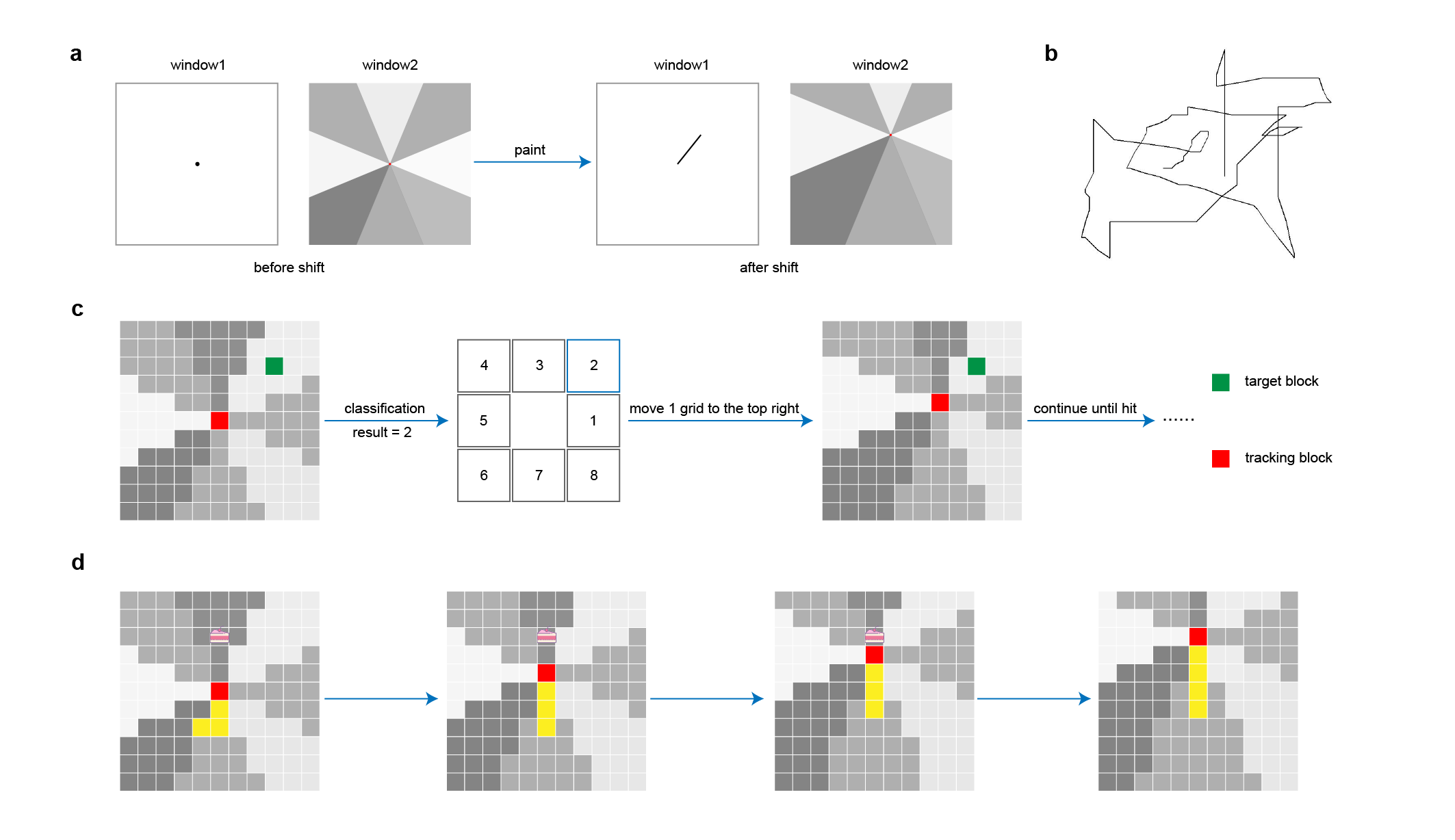}
  \caption{\textbf{Applications of visual tracking BCI. a,} Design of painting application. Window 1 for painting display, and window 2 for stimulation. 
    \textbf{b,} The painting result of one participant (Subject 3) depicts a dog. \textbf{c,} Schematic diagram of the visual tracking brain-computer 
    interface in discrete cases. \textbf{d,} Design and process of the snake game.}
\label{fig:5}
\end{figure*}
\subsection{Tracking task performance}
Based on the training session, we successfully achieved visual tracking in both fixed and random tracking tasks. \reffig{fig:4}{(\textbf{a})} illustrates the tracking trajectories of the fixed tracking task and random tracking task (example results come from Subject 13 and Subject 1, see \reftab{tab:1}). To evaluate the performance of tracking tasks, we calculated metrics such as time to target, success rate, and Fitt’s ITR. \reffig{fig:4}{(\textbf{b})} shows the distribution of time to target of the fixed tracking task for all participants. We separated the conditions into ‘inner’ (inner circle) and ‘outer’ (outer circle), and as expected, time to target of the inner circle was smaller than that of the outer circle (inner: mean = 3.53, s.d. = 0.74; outer: mean = 5.03, s.d. = 0.51). The average time to target of the random tracking task was 4.17 s, with a s.d. of 0.93 (as in \reffig{fig:4}{(\textbf{c})}). We then obtained Fitt’s ITR of the fixed tracking task and random tracking task for each participant (\reffig{fig:4}{(\textbf{d})}). The average of Fitt’s ITR in the fixed tracking task was 0.55 bps with a maximum of 0.68 bps (s.d. = 0.09), and the average of Fitt’s ITR in the random tracking task was 0.37 bps with a maximum of 0.54 bps (s.d. = 0.08). The results are comparable to related invasive studies, which typically yield a Fitt’s ITR of 0.4 bps \supercite{silversmith2021plug}. We attributed the lower Fitt’s ITR in the random tracking task to the fact that the distance to the random target for successful trials was close to the start position, while for farther targets, trials were more prone to failure. \reffig{fig:4}{(\textbf{e})} illustrates the success rate for the two tasks (fixed: mean = 92.6\%, s.d. = 7.1\%; random: mean = 84.8\%, s.d. = 12.9\%). Tracking trajectories for all participants in both tasks are depicted in \reffig{fig:S5}{} and \reffig{fig:S6}{}. And here, we showed the tracking trajectories of the representative subject for four specific target positions of fixed tracking task, including three tracking paths for each target position (Subject 3; \reffig{fig:4}{(\textbf{f})}, Left). Additionally, we displayed trajectories for the first 5 target positions of the random tracking task (\reffig{fig:4}{(\textbf{f})}, Right). For both the fixed and random tracking tasks, we recorded the videos to show the process (see \refvid{vid:1}\ and \refvid{vid:2}). The aforementioned results demonstrate our successful achievement of visual tracking and the attainment of a level of cursor movement control that matches invasive methods.

%% file: sections/5-applications.tex
\subsection{Applications}
\emph{Painting.} Expanding on the random tracking task, we designed a painting application for all participants to experience after the testing part. We positioned two windows on the screen: one for stimulation and the other for painting display (see \reffig{fig:5}{(\textbf{a})}). During the painting experience, we allowed participants to freely move their gaze to experience controlling the brush. The results were collected as a questionnaire asking participants, “Do you think the brush draws according to your will?” (Almost positive), and for participants who experienced BCI speller, we also asked, “Is the visual load of this application larger than BCI speller?” (All are negative; see \reftab{tab:1}). Fig. 5b shows the painting of one participant (Subject 3), depicting the image of a dog. The painting process for writing number 2 can be seen in \refvid{vid:3}.

\emph{Snake game.} To facilitate control for the snake game, we simplified the visual tracking BCI to a discrete case. Firstly, the stimulus region consists of stimulus blocks arranged radially around the central red tracking block. Secondly, the decoding algorithm only uses traditional classification methods. When the user fixates on the green target block (see \reffig{fig:5}{(\textbf{c})}), the system computes the classification result corresponding to the region where the target is being fixated. Subsequently, the red tracking block moves one grid space towards that region. This process continues until the red tracking block reaches and ‘hits’ the green target block. We used the discrete visual tracking BCI technology to design the snake game (as shown in \reffig{fig:5}{(\textbf{d})}; red block: head; yellow block: body). In the game, users control the snake’s movement up, down, left, and right by fixating on the corresponding regions. The regions in other directions do not possess movement control functions, allowing users to shift their gaze to observe the food's position without triggering unnecessary movement of the tracking block. The users need to control the snake to obtain food and avoid hitting the snake’s body or the edge of the screen. At the end, a score that stands for the length of the snake will be given. 17 participants played the snake game and got their scores (see \reffig{fig:S7}{}; maxima = 14). The gaming process was recorded in \refvid{vid:4}.

%% file: sections/6-discussion.tex
\begin{figure*}[t!]
    \centering
    \includegraphics[width=\textwidth]{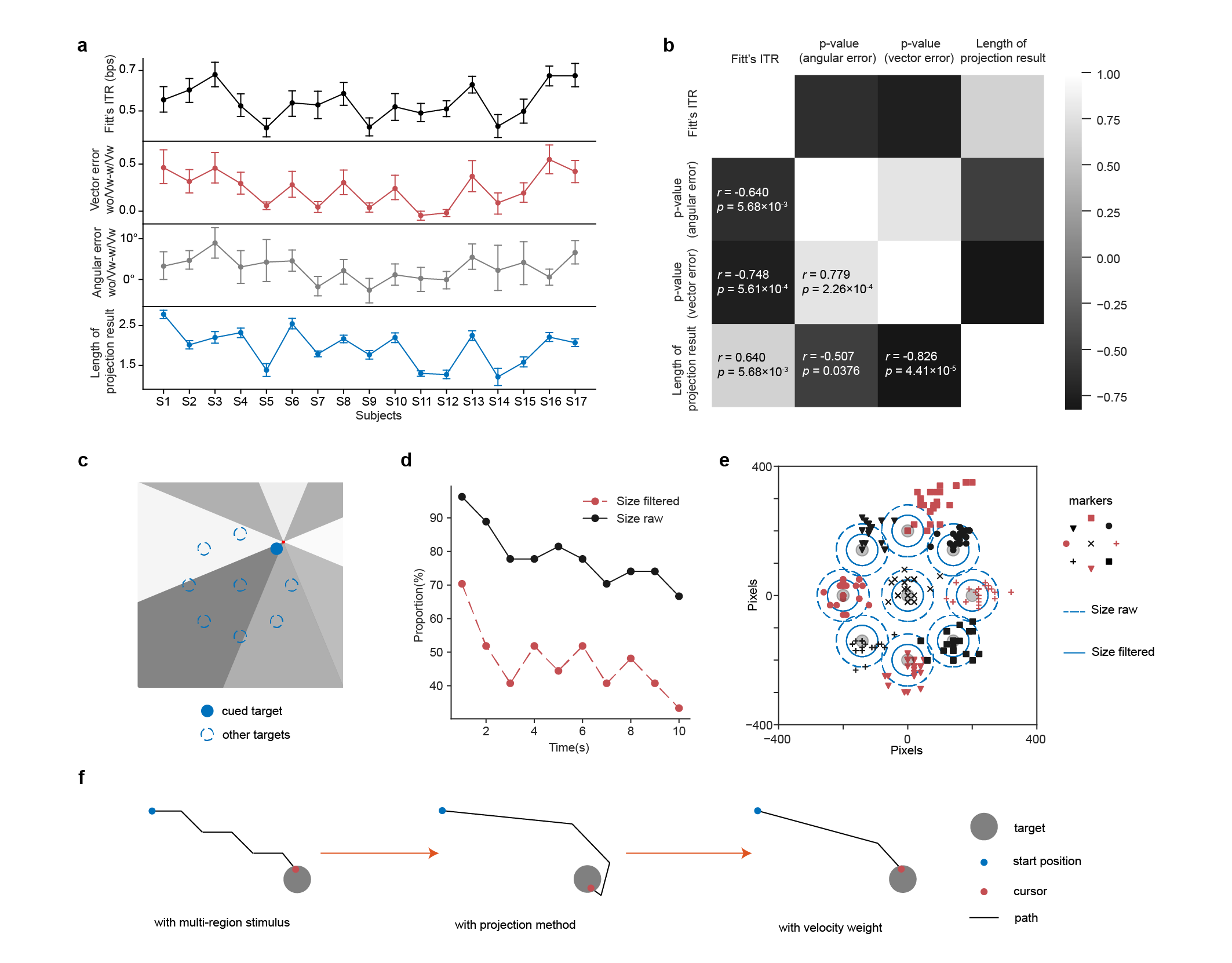}
    \caption{\textbf{Further evaluation and examination of visual tracking BCI. a,} Fitt’s ITR, vector error difference (wo/Vw-w/Vw), angular error difference (wo/Vw-w/Vw), and length of projection result of all participants. 
      Bars indicate mean ± 95\% CI. \textbf{b,} Spearman correlation coefficient matrix between Fitt’s ITR, p-value of vector error difference, p-value of angular error difference, and length of projection result (n = 17). 
      \textbf{c,} Schematic of the jitter inspection experiment. \textbf{d,} Proportion of jitter positions within ‘size filtered’ and within ‘size raw’ over time. 
      \textbf{e,} Map of jitter positions for each target. Different markers correspond to different targets. \textbf{f,} Schematic of path optimization.}
  \label{fig:6}
  \end{figure*}
\section{Discussion}
In this study, we proposed a novel spatial encoding stimulus paradigm and the corresponding velocity projection method to achieve continuous visual tracking based on noninvasive EEG. We achieved ideal results in velocity prediction and cursor control tasks across 17 subjects and successfully implemented various applications, showcasing the efficiency and practical utility of the visual tracking BCI. Furthermore, we have elucidated the mechanism of the stimulus paradigm through the lens of the receptive field, which signifies a novel approach to continuous decoding. In summary, this study transcends traditional discrete visual BCI and achieves continuous control based on a noninvasive method.
\subsection{Interpretability of BCI performance}
As we can notice, the continuous control performance experiences inter-subject variability. We have identified a metric for pre-estimating the performance of the participants in tracking tasks. The metric was defined as the length of the projection result from the data of training Stage II, which exhibited a robust correlation with other metrics of BCI performance. \reffig{fig:6}{(\textbf{a})} illustrates Fitt’s ITR, vector error difference, angular error difference, and the length of the projection result across participants, revealing a distinct correlation among them. Additionally, we assessed the significance levels (p-values) of vector error difference and angular error difference to replace the two differences and computed the Spearman correlation coefficients among the four metrics (see \reffig{fig:6}{(\textbf{b})}). The result shows a positive correlation between the length of the projection result and Fitt’s ITR (r = 0.640, P = $5.68 \times 10^{-3}$), suggesting that it can serve as an estimator for the task performance level of the participants. It is worth noting that the length of the projection result exhibits a negative correlation with both the p-value of vector error (r = -0.826, P = $4.41 \times 10^{-5}$) and angular error (r = -0.507, P = 0.0376), which illustrates that the length of projection result can predict the effectiveness of the corrected velocity weight. Based on our current observations, two situations lead to a longer projection result, both rooted in the subject's heightened responsiveness to stimulus encoding. Firstly, when the subject's attention field of vision is more focused, it is less influenced by the surrounding stimulus regions. This results in the subject primarily receiving stimulation from the region within the visual field, making the projection vector corresponding to that region dominant. Secondly, subjects with a broader field of vision are stimulated by more regions. However, owing to the spatial encoding characteristics, the sum of vectors (projection result) accumulates in the target direction and nullifies in other directions. This cumulative effect ultimately yields a longer projection result.
\subsection{Gaze fixation and path optimization}
For human-computer interaction similar to cursor control, in addition to the requirement of reaching the target, the accuracy of the movement path largely determines the efficiency and functionality of the system. As shown in \reffig{fig:6}{(\textbf{f})}, path optimization is essential for the performance of visual tracking BCI. Utilizing solely the multi-region stimulation paradigm can only facilitate discrete visual tracking. However, by incorporating the projection method, continuous visual tracking can be achieved. Furthermore, in combination with velocity weight, velocity can be refined for enhanced accuracy and reduced tracking time. However, besides achieving continuous smooth tracking, knowing when to stop is also crucial to enabling the functionality of selection. Therefore, we investigated the issue of cursor jitter to determine the range and time of the users' gaze fixation. To detect the jitter of the tracking point (cursor) in the visual tracking BCI, we conducted an experiment in which users were instructed to fixate on the cued target (including 9 target positions; \reffig{fig:6}{(\textbf{c})}) while recording the jitter positions within the subsequent 10 seconds. Subject 6 participated in the experiment, which was repeated 3 times for each target position. For the evaluation of jitter intensity, we employed the target recognition range from eye-tracking technology as a reference \supercite{feit2017toward}. This range measured 1.9 x 2.35 cm (size filtered; square diagonal length as circular diameter) when utilizing a filter and 3.28 x 3.78 cm (size raw) when not. Then, we found that this technology can achieve the maintenance of gaze fixation within a certain time and range. \reffig{fig:6}{(\textbf{d})} shows the variation of the proportion of jitter positions within ‘size filtered’ and within ‘size raw’ over time (size raw: 1s, 96.3\%; size filtered: 1s, 70.4\%). And we presented all jitter positions for each target in \reffig{fig:6}{(\textbf{e})}. In the context of the tracking task, we define ‘hit’ as the moment that the cursor makes contact with the target. However, during that instant, the cursor's velocity is not zero due to the 1-second time step employed. As a result, we ascertained whether the cursor continues to maintain ‘hit’ the target after a certain duration (see \reffig{fig:S8}{}; 0.2s, 83.5\%).
\subsection{Comparison and future improvement}
There are currently two established BCI methods for cursor movement tasks. One involves an invasive spike-based BCI that utilizes Kalman filter-based algorithms \supercite{silversmith2021plug,gilja2012high,hochberg2012reach}. The other is a non-invasive motor imagery BCI, primarily employing EEG signals and the velocity algorithm proposed by Wolaw et al.\supercite{wolpaw2004control}. In our study, we achieved an average Fitt’s ITR of 0.55 bps in the fixed visual tracking task, often referred to as a center-out task in other research. These results compare favorably to some spike-based BCIs, which have reported Fitt's ITR of 0.4 bps\supercite{silversmith2021plug}, although they fall below the highest recorded rates for individuals ranging from 1.4 to 4.2 bps\supercite{pandarinath2017high}. In comparison to non-invasive BCIs, our results surpass the 0.29 bps calculated from the presented data of a 2D BCI task using Wolpaw's method\supercite{felton2009evaluation}.

Future improvements should focus on stimulus design and algorithm optimization, real-time asynchronous system construction, and click function implementation. The dynamic sequences we employed here are the WN sequences; however, it's worth noting that other sequences, such as sinusoidal sequences or m-sequences\supercite{bin2011high}, can also be tested. Importantly, stimulus sequences from different regions may not necessarily have an equal impact on participants\supercite{gulbinaite2019attention}. This can pose challenges when tracking targets within specific directional ranges. In this study, we did not extensively investigate the varying effects of different stimulus sequences on participants. However, exploring these differences could potentially provide a path to enhance BCI performance\supercite{ye2022multisymbol}. From a different perspective, non-equivalent stimulus sequences, such as sinusoidal sequences with different frequencies, may hold greater potential in certain scenarios with preferences for specific directions. In this study, we employed Task-Related Component Analysis (TRCA) as the template matching method. However, other classification-based algorithms such as Task-Discriminant Component Analysis (TDCA)\supercite{liu2021improving}, or Filter Bank Canonical Correlation Analysis (FBCCA)\supercite{chen2015high} for SSVEP are also possible. Furthermore, it's important to note that we utilized only one type of target size, and the training of velocity weight relied solely on a linear method. Consequently, we believe that there is substantial potential for enhancing the task performance of visual tracking. We aspire for future work to employ shorter time steps or build asynchronous systems\supercite{chen2019communication} to achieve real-time visual tracking. Alternatively, a sliding window method can be used to collect 1 s of data every 0.1 s for analysis to improve the temporal resolution of decoding. For the click function, we aim to realize it through dwell time\supercite{carter2020best} or in combination with the brain-computer interface of motor imagery\supercite{li2010eeg}. Combining these improvements, this technology will ultimately achieve a more efficient, comfortable, and natural human-computer interaction effect.

%% file: sections/si.tex
\newpage
\centerline{\textbf{Supplementary Information}}
\renewcommand\thefigure{S\arabic{figure}}
\renewcommand\thetable{S\arabic{table}}
\setcounter{figure}{0}
\begin{figure}[H]
  \centering
  \includegraphics[width=\textwidth]{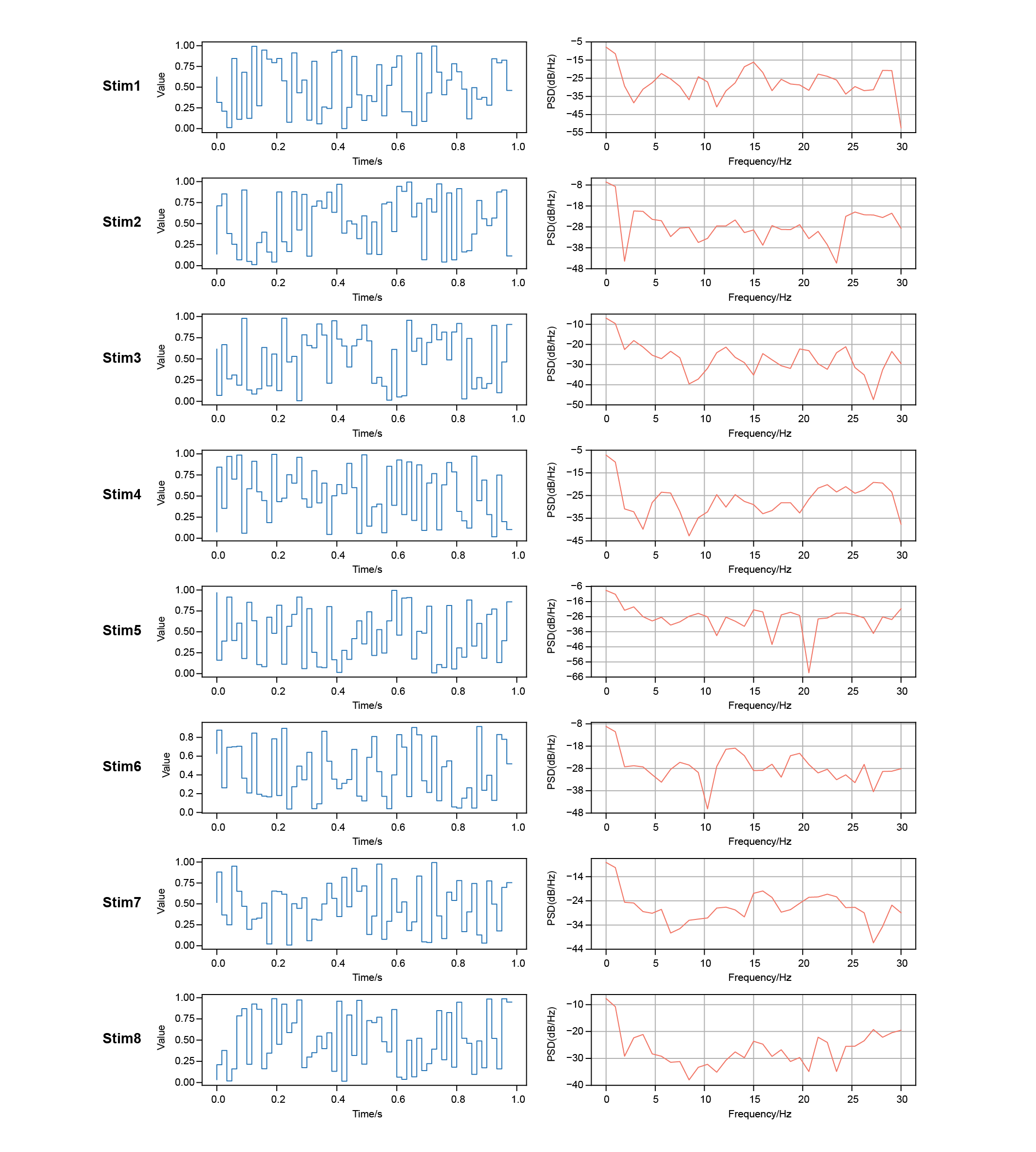}
  \caption{\textbf{Characteristics of WN sequences.} The waveforms and power spectral density (PSD) of the 8 WN sequences used in practice are displayed here, illustrating their uniform distribution between 0 and 1, as well as the 0-30 Hz spectrum due to the 60 Hz screen refresh rate limitation.}
\label{fig:S1}
\end{figure}
\clearpage
\begin{figure}[H]
  \centering
  \includegraphics[width=\textwidth]{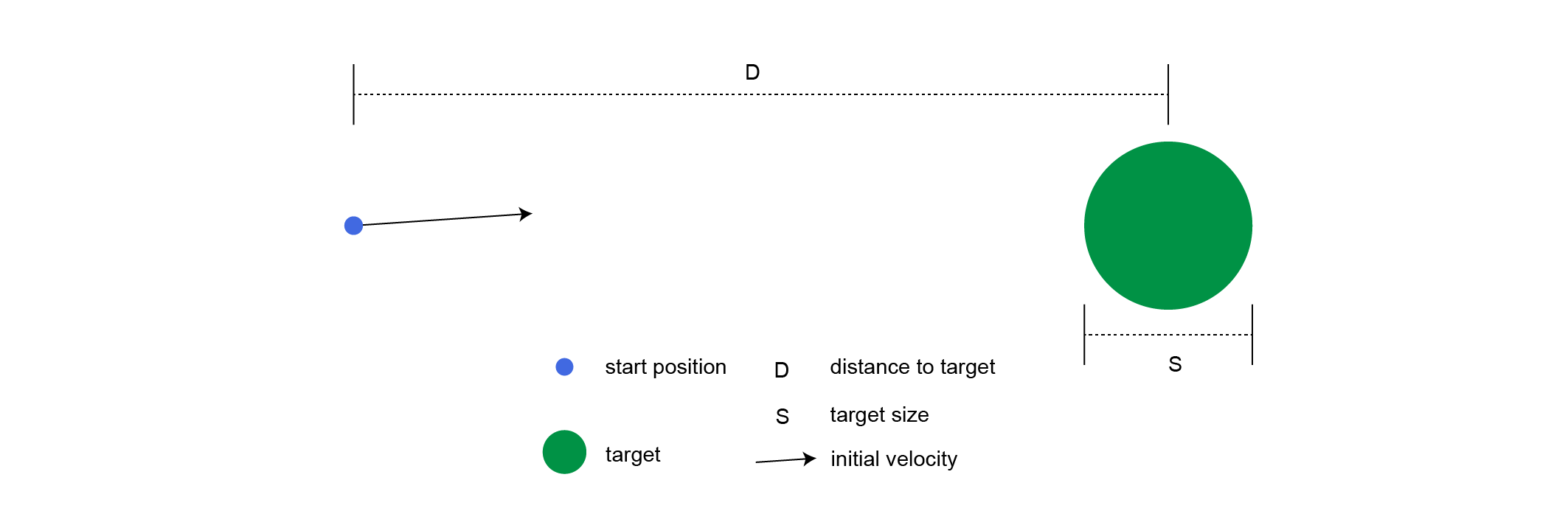}
  \caption{\textbf{Schematic diagram of Fitt’s ITR parameters.} In this context, ‘distance to target’ pertains to the distance from the starting position of a tracking task to the center of the target, while ‘target size’ denotes the diameter of the green circle.}
\label{fig:S2}
\end{figure}
\clearpage
\begin{figure}[H]
  \centering
  \includegraphics[width=\textwidth]{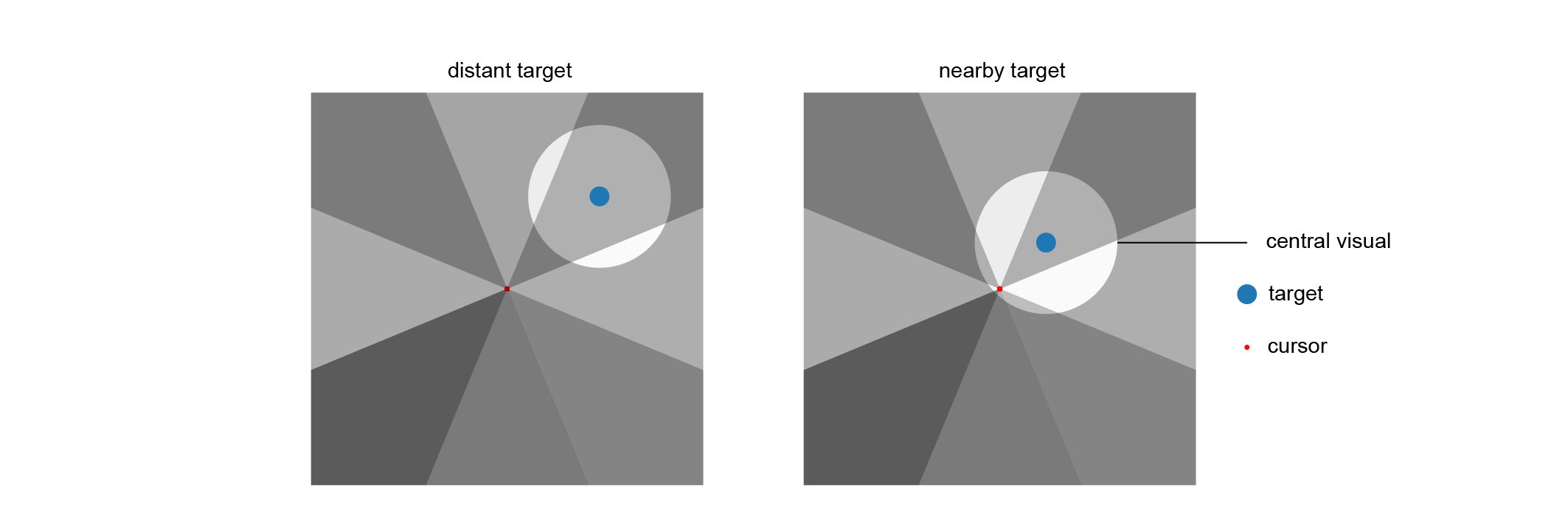}
  \caption{\textbf{Schematic diagram of stimulation mechanism.} Left: the situation of a distant target, the proportion of surrounding regions’ stimulus in the central visual is smaller; Right: the situation of a nearby target, the proportion of surrounding regions’ stimulus is larger.}
\label{fig:S3}
\end{figure}
\clearpage
\begin{figure}[H]
  \centering
  \includegraphics[width=\textwidth]{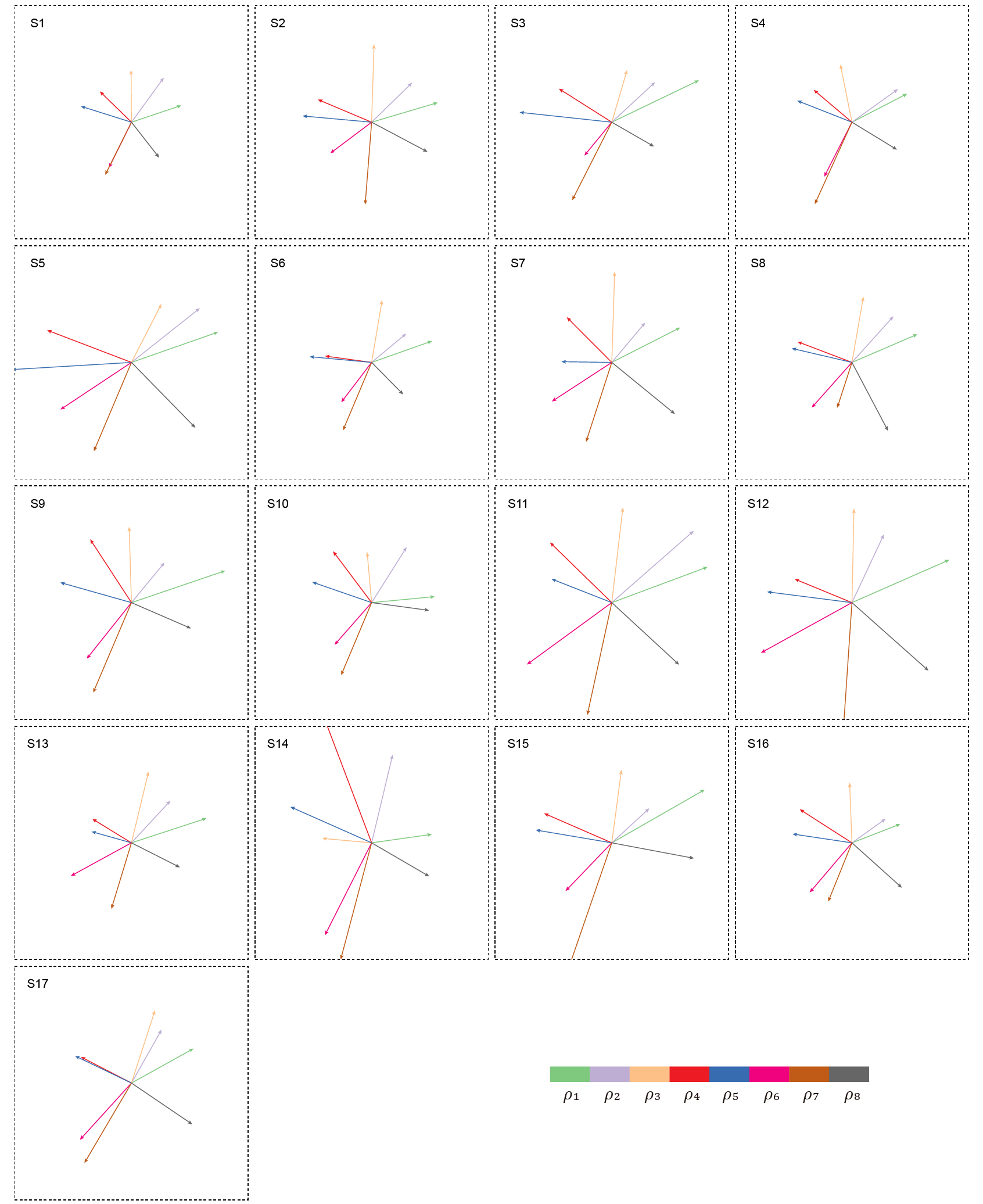}
  \caption{\textbf{Velocity weight of all participants.} Here, we used the form of vectors to show the velocity weight. For each $\rho$, there is a velocity weight with a corresponding color (in color bars). It's worth noting that the velocity weight does not align with the central direction of the $\rho$-corresponding region; instead, it serves a corrective role.}
\label{fig:S4}
\end{figure}

\clearpage
\begin{figure}[H]
  \centering
  \includegraphics[width=\textwidth]{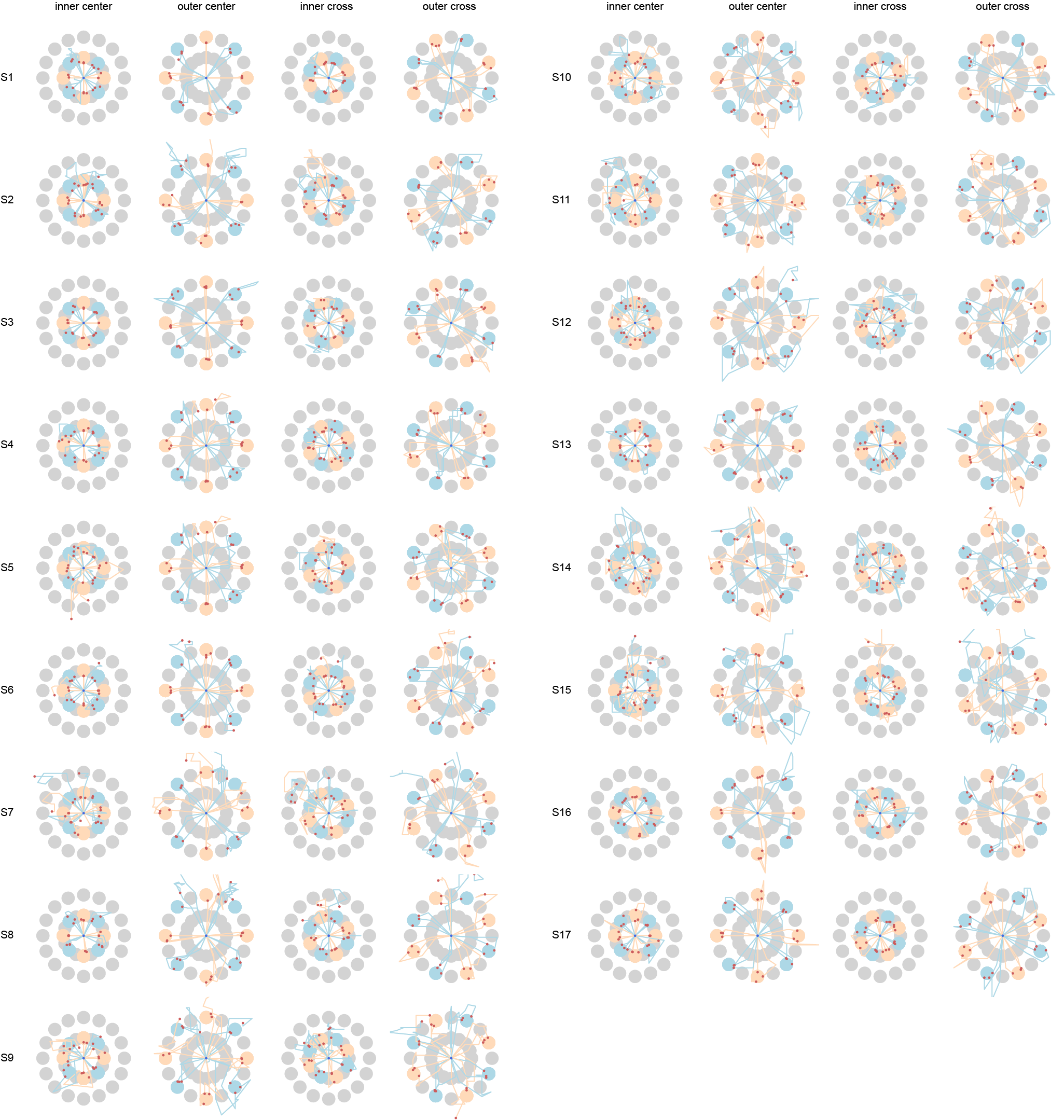}
  \caption{\textbf{Tracking trajectories for fixed tracking task of all participants.} Here we plotted all tracking trajectories in the fixed tracking task with four conditions (inner center, outer center, inner cross, and outer cross; see Results for Effect of velocity prediction and velocity weight). We used two different colors in each figure to make the trajectories more distinguishable.}
\label{fig:S5}
\end{figure}
\clearpage
\begin{figure}[H]
  \centering
  \includegraphics[width=\textwidth]{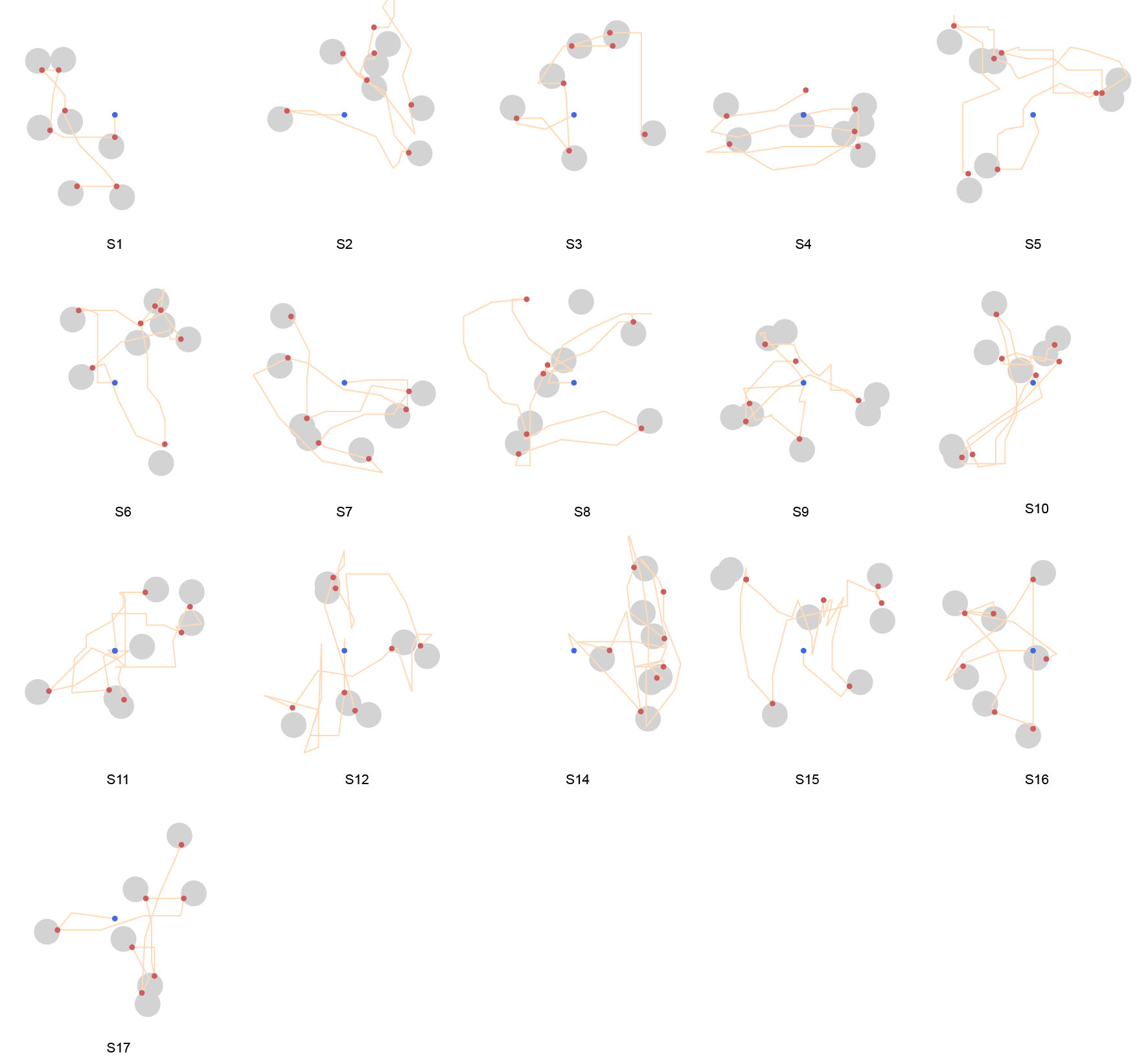}
  \caption{\textbf{Tracking trajectories for random tracking task of all participants.} Here, we only showed the first 7 targets in the task. The data of S13 were not recorded correctly in the experiment.}
\label{fig:S6}
\end{figure}
\clearpage
\begin{figure}[H]
  \centering
  \includegraphics[width=0.8\textwidth]{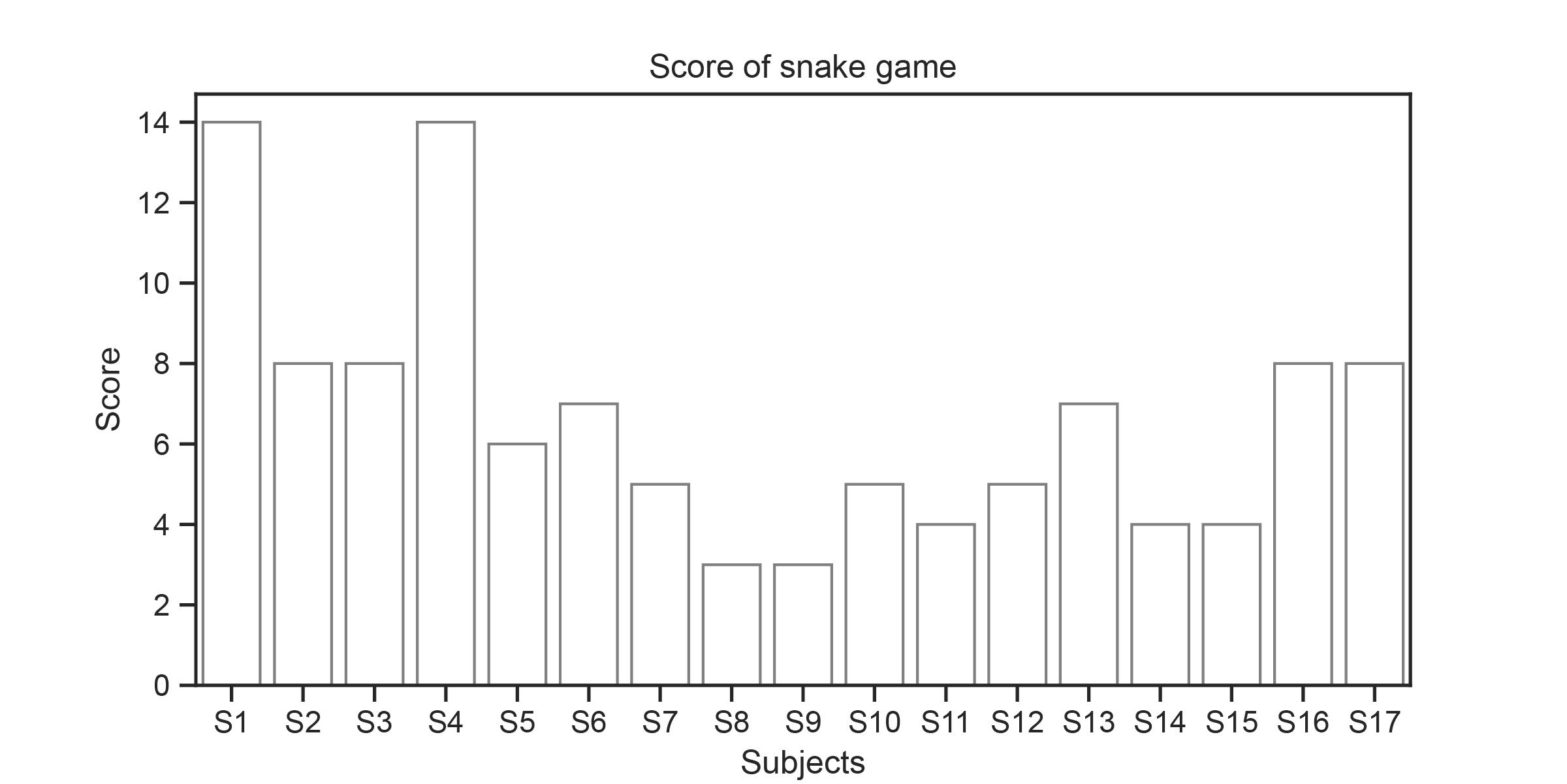}
  \caption{\textbf{Score of snake game.} The conditions with a score less than 5 happened when the participants were too tired to continue or it was difficult for the participant to get a higher score. Maxima = 14, Minima = 3.}
\label{fig:S7}
\end{figure}
\clearpage
\begin{figure}[H]
  \centering
  \includegraphics[width=0.5\textwidth]{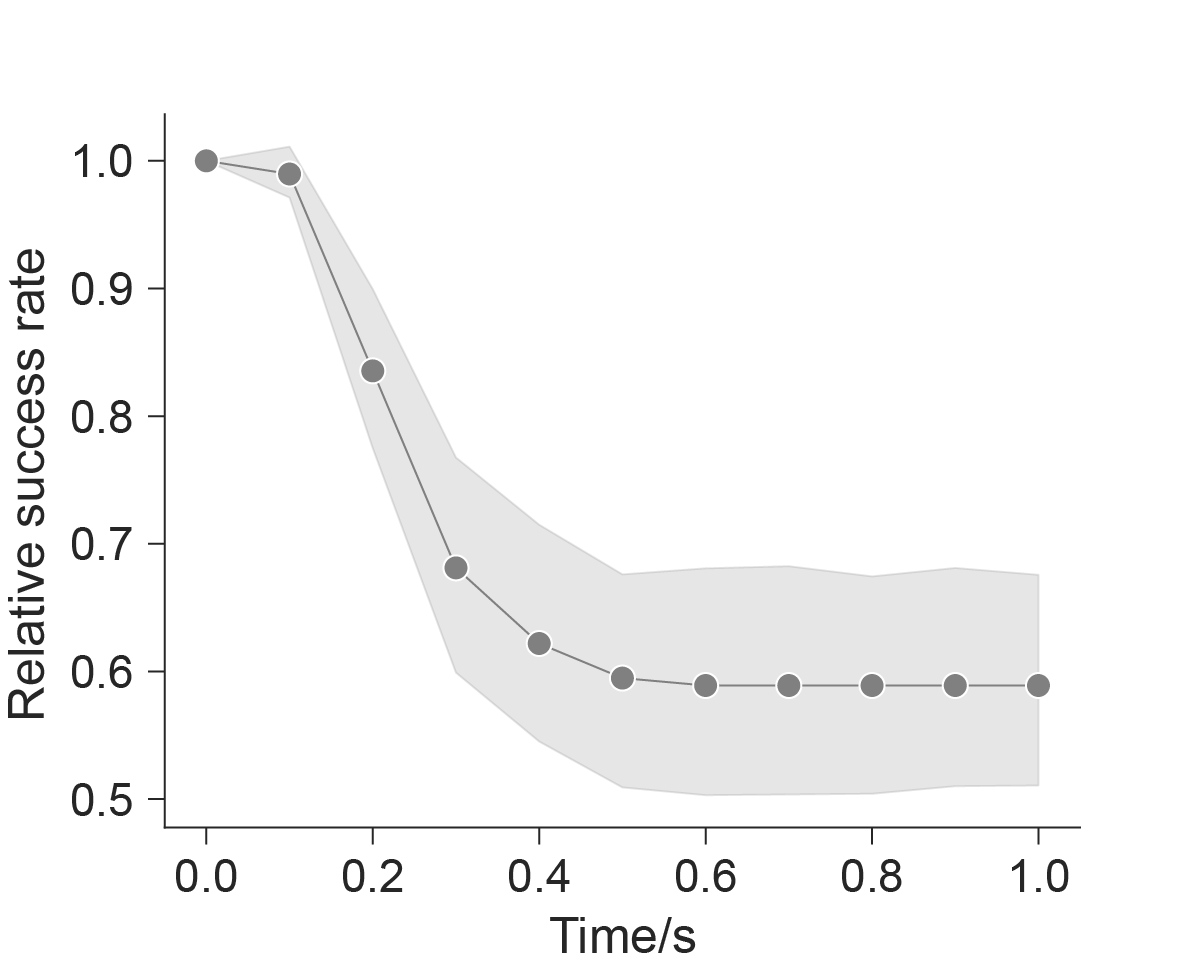}
  \caption{\textbf{Relative success rate after a period of time.} The figure shows the relationship between the relative success rate (success rate after a period of time ratios to the initial success rate) and time after the first ‘hit’ moment. The value of the relative success rate for 0.2 s is 83.5\%.}
\label{fig:S8}
\end{figure}

\clearpage
\begin{table}[H]
\caption{\textbf{Participants information and results of the questionnaire.} Table includes information of participants’ gender and BCI experience. Question 1 is “Do you think the brush draws according to your will?” for all participants, and Question 2 is “Is the visual load of this application larger than BCI speller?” for participants with BCI experience.}
\centering
\begin{tabular}{|c|c|c|c|c|}
\hline
\textbf{Subjects} & \textbf{Gender} & \textbf{BCI experience} & \textbf{Question1} & \textbf{Question2} \\ \hline
S1                & male            & experienced             & yes                & no                 \\ \hline
S2                & male            & naive                   & yes                &                    \\ \hline
S3                & male            & experienced             & yes                & no                 \\ \hline
S4                & male            & experienced             & yes                & no                 \\ \hline
S5                & male            & naive                   & no                 &                    \\ \hline
S6                & male            & naive                   & yes                &                    \\ \hline
S7                & female          & naive                   & yes                &                    \\ \hline
S8                & male            & experienced             & yes                & no                 \\ \hline
S9                & male            & experienced             & yes                & no                 \\ \hline
S10               & male            & naive                   & yes                &                    \\ \hline
S11               & male            & naive                   & yes                &                    \\ \hline
S12               & female          & naive                   & yes                &                    \\ \hline
S13               & female          & experienced             & yes                & no                 \\ \hline
S14               & male            & naive                   & yes                &                    \\ \hline
S15               & female          & naive                   & yes                &                    \\ \hline
S16               & female          & experienced             & yes                & no                 \\ \hline
S17               & female          & naive                   & yes                &                    \\ \hline
\end{tabular}
\label{tab:1}
\end{table}

\clearpage
\begin{video}[H]
\caption{\textbf{Example of the fixed tracking task.} This video shows performance of one subject during fixed tracking task. Visit \href{https://peanut-1210.github.io/Oser/}{https://peanut-1210.github.io/Oser/}}
\label{vid:1}
\end{video}
\begin{video}[H]
\caption{\textbf{Example of the random tracking task.} This video shows performance of one subject during random tracking task. Visit \href{https://peanut-1210.github.io/Oser/}{https://peanut-1210.github.io/Oser/}}
\label{vid:2}
\end{video}
\begin{video}[H]
\caption{\textbf{Example of the painting application.} In this video, we asked the subject to write number 2. Here you can see how she writes. Visit \href{https://peanut-1210.github.io/Oser/}{https://peanut-1210.github.io/Oser/}}
\label{vid:3}
\end{video}
\begin{video}[H]
\caption{\textbf{Example of the snake game application.} This video shows the game process of one subject. You can clearly see how the snake moves and eats the food. Visit \href{https://peanut-1210.github.io/Oser/}{https://peanut-1210.github.io/Oser/}, and Supplementary Video.5 in the website shows the experiments scene.}
\label{vid:4}
\end{video}